\documentstyle[preprint,aps]{revtex}

\draft

\begin{document}

\title{Study of chaos in hamiltonian systems via convergent normal forms}

\date{\today }

\author{Werner M. Vieira$^{\dag}$, Alfredo M. O. de Almeida$^{\ddag}$}

\maketitle

\begin{abstract}

We use Moser's normal forms to study chaotic motion in two-degree hamiltonian
 systems near a saddle point. Besides being convergent, they provide a
 suitable description of the cylindrical topology of the chaotic flow in that
 vicinity. Both aspects combined allowed a precise computation of the
 homoclinic interaction of stable and unstable manifolds in the full phase 
space, rather than just the Poincar\'e section. The formalism was applied to
 the H\'enon-Heiles hamiltonian, producing strong evidence that the region of
 convergence of these normal forms extends over that originally established
 by
 Moser.

\end{abstract}
PACS index:05.45.+b\\

\footnoterule

{\footnotesize{$^{\dag}$Departamento de Matem\'atica Aplicada, Universidade 
Estadual de Campinas, CP 6065, 13083-970 Campinas, SP, Brazil. e-mail: 
vieira@ime.unicamp.br

$^{\ddag}$Centro Brasileiro de Pesquisas F\'{\i}sicas, Rua Xavier Sigaud,
 150, 2290-180 Rio de Janeiro, RJ, Brazil.}}

\newpage

\section{INTRODUCTION}\label{sec.1}

Normal Forms (NF) are among the successful methods for either  analytic 
or numeric studies of  dynamical systems; by performing a suitable
 coordinate transformation, we eventually obtain a more simple or
 dynamically ``transparent'' version of the original system. This
 approach can be formulated  either for generic systems including
 two-dimensional maps~\cite{1} or for hamiltonian systems~\cite{2}
 and conservative maps~\cite{3}. So, even when a given NF does not
 converge, due to small denominators or exact ressonances, it is of
 importance for numerical purposes. This is just what occurs when the NF is
 obtained around a stable point or orbit. Then, in spite of the well known 
divergence in this case, a truncation allows us to follow the motion for a 
long time with great precision. In fact, the nonconvergent case is the most
 considered in the literature~\cite{20,5,19}.

The present work concerns the NF around a {\it{unstable}} point or orbit of
 a conservative system. For that, Moser demonstrated their convergence for
 both maps~\cite{7} and hamiltonian systems~\cite{8}. Although convergent, 
this case did not receive much attention until recently, presumably because
 the particle remains a very short time in that region. Nevertheless, we will
 see that the Moser normal forms (MNF) are both convergent and a powerfull
 tool for the search for the basic structures of the chaotic motion rather
 than just following a specific orbit for a long time.

The usefulness of the MNF for the study of conservative chaotic maps is
 already known in the literature. It allowed precise analytical computations
 of homoclinic points and the periodic points with long period, which
 accumulate in the homoclinic ones~\cite{11,12}. Additional good numerical
results were obtained even if small dissipative perturbations were added 
~\cite{13}.

Area-preserving maps are usually only simplified reductions (appropriate
 Poin\-car\'e sections) of autonomous hamiltonian systems of two degrees
 of freedom. In particular, the very complex two-dimensional homoclinic 
tangle~\cite{10} is already a reduction of the much more involved chaotic
 motion in the full phase space. However, extending the use of the MNF to
 the hamiltonian case allows us to study directly the proper four-dimensional
 chaotic flow. This has previously been attempted in the literature~\cite{15}
, but without taking full advantage of the method that we shall develop here.
 Another use of MNFs is found in~\cite{50} and takes advantage of its 
convergence to ascertain stability transitions of families of periodic orbits 
near hamiltonians' saddle points.

In section \ref{sec.2} we will develop the MNF approach for the case of a
 generic autonomous hamiltonian of two degrees of freedom around a saddle
 point, encompassed by Moser's convergence proof. In that vicinity, the 
flow's topology is {\it{cylindrical}} rather than toroidal, in the case of
 a chaotic regime~\cite{17}. Firstly, we construct the relations linking 
the original system to the corresponding more ``transparent'' normalized 
system. This is done through a near identity polynomial coordinate 
transformation. We will see that, besides convergent, that transformation 
also reveals in a  natural way, the cylindrical character of the topology. 
Both features turn the MNF into a powerfull tool. So, it was possible to 
compute, precisely for the first time, the {\it{continuous}} structures in
 the full phase space, underlying the homoclinic tangle in a Poincar\'e 
section: the (un)stable manifolds which originate at the saddle point and
 at each neighbouring unstable periodic orbit, the homoclinic orbits 
associated with the latter and the periodic orbits with long period which 
accumulate on the homoclinic orbits. In section \ref{sec.3} we obtain the 
recurrence relations for the coefficients involved in the theory.

In sections \ref{sec.4} and \ref{sec.5} we apply the formalism to the 
specific case of the H\'enon-Heiles hamiltonian. The numerical results 
exhibited in section \ref{sec.5} fully confirmed the expectations about
 the MNF as a tool for the study and characterization of chaotic motions. 
Moreover, they also point to some kind of extension of the region of 
convergence initially assumed for Moser's theorem. In fact, this issue is
 just being considered by the authors presently.

Finally, in section \ref{sec.6} we summarize the results and possible
 extensions of the present work.

\section{MOSER'S NORMAL FORM}\label{sec.2}

It is essential that the hamiltonian be in the complexified form, 
for the method's implementation:
\begin{equation}
h(x_{1},x_{3},x_{2},x_{4})\;=\;\lambda_{1}x_{1}x_{3}+\lambda_{2}x_{2}x_{4}
+\sum_{\ell=3}^{\infty}H(\underline{\ell})\,\underline{x}^{\underline{\ell
}}{\mbox{\hspace{0.4cm}.}}\label{eq.1}
\end{equation}
Here, the origin is assumed to be a saddle point and we use the notation: 
$\underline{\ell}=(\ell_{1},\ell_{3},\ell_{2},\ell_{4})$, 
$\underline{x}=(x_{1},x_{3},x_{2},x_{4})$, 
$\ell=\ell_{1}+\ell_{3}+\ell_{2}+\ell_{4}$ and 
$\underline{x}^{\underline{\ell}}=x_{1}^{\ell_{1}}x_{3}^{\ell_{3}}x_{2}
^{\ell_{2}}x_{4}^{\ell_{4}}$. The positions are $x_{1}$ and $x_{2}$ and
 the conjugate momenta are, respectively, $x_{3}$ and $x_{4}$. The
 eigenvalues of the system's linear part are $\lambda_{1}=i\omega$ and 
$\lambda_{2}=-\lambda$, with $\omega$ and $\lambda$ real (the other two
 being of course $-i\omega$ and $\lambda$). $H(\underline{\ell})$ is the 
coefficient of $\underline{x}^{\underline{\ell}}$ and $\ell$ is its order.

The usual noncomplexified form of the quadratic part $h_{2}$ of (\ref{eq.1}), 
around the saddle point, is
\begin{equation}
h_{2}(q_{1},q_{2},p_{1},p_{2})\;=\;\frac{\lambda}{2}(p_{2}^{2}-q_{2}^{2})
+\frac{\omega}{2}(p_{1}^{2}+q_{1}^{2}){\mbox{\hspace{0.4cm},}}\label{eq.1.1}
\end{equation}
both sets of coordinates being linked by the symplectic transformation
\begin{equation}
\left\{
\begin{array}{l}
q_{1}\;=\;\frac{1}{\sqrt{2}}(x_{1}+ix_{3})\\
p_{1}\;=\;\frac{1}{\sqrt{2}}(ix_{1}+x_{3})\\
q_{2}\;=\;\frac{1}{\sqrt{2}}(x_{2}+x_{4})\\
p_{2}\;=\;\frac{1}{\sqrt{2}}(-x_{2}+x_{4}){\mbox{\hspace{0.4cm}.}}\label
{eq.1.2}
\end{array}
\right.
\end{equation}

The Hamilton equations coming from (\ref{eq.1}) are:
\begin{equation}
\left\{
\begin{array}{l}
\dot{x}_{1}\;=\;\lambda_{1}x_{1}+\sum_{\ell=3}^{\infty}\ell_{3}\,H(
\underline{\ell})\,x_{1}^{\ell_{1}}x_{3}^{\ell_{3}-1}x_{2}^{\ell_{2}}x
_{4}^{\ell_{4}} \\
\dot{x}_{3}\;=\;-\lambda_{1}x_{3}-\sum_{\ell=3}^{\infty}\ell_{1}\,H
(\underline{\ell})\,x_{1}^{\ell_{1}-1}x_{3}^{\ell_{3}}x_{2}^{\ell_{2}}
x_{4}^{\ell_{4}} \\
\dot{x}_{2}\;=\;\lambda_{2}x_{2}+\sum_{\ell=3}^{\infty}\ell_{4}\,H(
\underline{\ell})\,x_{1}^{\ell_{1}}x_{3}^{\ell_{3}}x_{2}^{\ell_{2}}x_{4}
^{\ell
_{4}-1} \\
\dot{x}_{4}\;=\;-\lambda_{2}x_{4}-\sum_{\ell=3}^{\infty}\ell_{2}\,H(
\underline{\ell})\,x_{1}^{\ell_{1}}x_{3}^{\ell_{3}}x_{2}^{\ell_{2}-1}
x_{4}^{\ell_{4}}{\mbox{\hspace{0.4cm}.}}\label{eq.2}
\end{array}
\right.
\end{equation}

Moser's theorem assures that there exists a near-identity polynomial 
coordinate transformation, convergent in a neighbourhood of the origin,
\begin{equation}
x_{i}\;=\;y_{i}+\sum_{\ell=2}^{\infty}X(i,\underline{\ell})\,\underline{
y}^{\underline{\ell}}{\mbox{\hspace{0.4cm},\hspace{1.3cm}$i\;=\;1,2,3,4$
\hspace{0.4cm},}}\label{eq.3}
\end{equation}
such that under it the original system (\ref{eq.2}) takes the 
{\it{normalized}} form
\begin{equation}
\left\{
\begin{array}{l}
\dot{y}_{1}\;=\;F(y_{1}y_{3}\,,\,y_{2}y_{4})\,y_{1} \\
\dot{y}_{3}\;=\;-F(y_{1}y_{3}\,,\,y_{2}y_{4})\,y_{3} \\
\dot{y}_{2}\;=\;G(y_{1}y_{3}\,,\,y_{2}y_{4})\,y_{2} \\
\dot{y}_{4}\;=\;-G(y_{1}y_{3}\,,\,y_{2}y_{4})\,y_{4}{\mbox{\hspace{0.4cm}
,}}\label{eq.4}
\end{array}
\right.
\end{equation}
the functions $F$ and $G$ depending as indicated only on the products
 $y_{1}y_{3}$ and $y_{2}y_{4}$. In fact, the normal system (\ref{eq.4}) is
 also hamiltonian, with Hamilton function $k(\underline{y})$ given by
\begin{equation}
k(y_{1},y_{3},y_{2},y_{4})\;=\;\lambda_{1}y_{1}y_{3}+\lambda_{2}y_{2}y_{4}+
\sum_{m=2}^{\infty}K(2\underline{m})\,(y_{1}y_{3})^{m_{1}}(y_{2}y_{4})^{m
_{2}}{\mbox{\hspace{0.4cm},}}\label{eq.7}
\end{equation}
where $K(2\underline{m})=K(2m_{1},2m_{2})$ are the expansion's coefficients.

It is straightforward to see that the two constants of motion of (\ref{eq.4}) 
are just the products
\begin{equation}
y_{1}y_{3}\;=\;{\mbox{$c_{1}$\hspace{0.4cm},}}\label{eq.5}
\end{equation}
\begin{equation}
y_{2}y_{4}\;=\;{\mbox{$c_{2}$\hspace{0.4cm},}}\label{eq.6}
\end{equation}
in terms of which the system can immediately be integrated:
\begin{equation}
\left\{
\begin{array}{l}
y_{1}(t)\;=\;y_{1}(0)\,\exp{(t\,F)} \\
y_{3}(t)\;=\;y_{3}(0)\,\exp{(-\,t\,F)} \\
y_{2}(t)\;=\;y_{2}(0)\,\exp{(t\,G)} \\
y_{4}(t)\;=\;y_{4}(0)\,\exp{(-\,t\,G)}{\mbox{\hspace{0.4cm},}}\label{eq.8}
\end{array}
\right.
\end{equation}
with $F$ and $G$ given by
\begin{equation}
\left\{
\begin{array}{l}
F(y_{1}y_{3},y_{2}y_{4})\;=\;\lambda_{1}+\sum_{m=2}^{\infty}m_{1}\,K(2
\underline{m})\,(y_{1}y_{3})^{m_{1}-1}(y_{2}y_{4})^{m_{2}} \\
G(y_{1}y_{3},y_{2}y_{4})\;=\;\lambda_{2}+\sum_{m=2}^{\infty}m_{2}\,K(2
\underline{m})\,(y_{1}y_{3})^{m_{1}}(y_{2}y_{4})^{m_{2}-1}{\mbox{\hspace
{0.4cm}.}}\label{eq.9}
\end{array}
\right.
\end{equation}

It is important to note that the coordinate transformation \ref{eq.3} need
 not to be canonical, even though the transformed system is also
 hamiltonian. The cylindrical topology of the flow can be made explicit. 
As the original coordinates $q_{1},q_{2},p_{1},p_{2}$ are real, we see 
from (\ref{eq.1.2}) that $x_{2}$ and $x_{4}$ are real and $x_{1}$ and 
$x_{3}$ are complex, there existing the following relation between $x_{1}$
 and $x_{3}$:
\begin{equation}
x_{3}\;=\;-i\,\overline{x}_{1}{\mbox{\hspace{0.4cm}.}}\label{eq.10}
\end{equation}
The overline symbol above indicates the complex conjugate operation. 
Now, the first order truncation of (\ref{eq.3}) permits us to extend the 
above conclusions to the corresponding coordinates $\underline{y}$, namely 
$y_{2}$ and $y_{4}$ are real and $y_{1}$ and $y_{3}$ are complex satisfying
 the relations
\begin{equation}
y_{3}\;=\;-i\,\overline{y}_{1}{\mbox{\hspace{0.4cm},\hspace{0.4cm}}}|y_{1}
|=|y_{3}|\equiv\rho{\mbox{\hspace{0.4cm},}}\label{eq.11}
\end{equation}
\begin{equation}
y_{1}y_{3}\;=\;-i\,\rho^{2}{\mbox{\hspace{0.4cm},}}\label{eq.12}
\end{equation}
$\rho$ being a new constant of motion. Then, it follows from (\ref{eq.8}) 
that $F$ is pure imaginary and $G$ is real. If we define (remembering that
 $\lambda_{1}=i\omega$ and $\lambda_{2}=-\lambda$)
\begin{equation}
\left\{
\begin{array}{l}
\Omega\;\equiv\;F/i\;=\;\omega+\delta\omega\\
\Lambda\;\equiv\;G\;=\;\lambda+\delta\lambda{\mbox{\hspace{0.4cm},}}\label
{eq.12.1}
\end{array}
\right.
\end{equation}
and
\begin{equation}
\left\{
\begin{array}{l}
\delta\omega\;\equiv\;-i\,\sum_{m=2}^{\infty}m_{1}\,K(2\underline{m})\,(-i
\rho^{2})^{m_{1}-1}(\epsilon)^{m_{2}}\\
\delta\lambda\;\equiv\;\sum_{m=2}^{\infty}m_{2}\,K(2\underline{m})\,(-i
\rho^{2})^{m_{1}}(\epsilon)^{m_{2}-1} {\mbox{\hspace{0.4cm},}}\label{eq.13}
\end{array}
\right.
\end{equation}
we can rewrite (\ref{eq.8}) as
\begin{equation}
\left\{
\begin{array}{l}
y_{1}(t)\;=\;\rho\,\exp{[i(\theta_{0}+\Omega\,t)]} \\
y_{3}(t)\;=\;\rho\,\exp{[-i(\theta_{0}+\frac{\pi}{2}+\Omega\,t)]} \\
y_{2}(t)\;=\;y_{20}\,\exp{[\Lambda\,t]} \\
y_{4}(t)\;=\;y_{40}\,\exp{[-\Lambda\,t]}{\mbox{\hspace{0.4cm},\hspace{0.4cm}
$y_{2}y_{4}\;=\;\epsilon$\hspace{0.4cm},}}\label{eq.14}
\end{array}
\right.
\end{equation}
so that $\epsilon$ is the second constant of motion. The convergence of the 
MNF guarantees that $\rho$ and $\epsilon$ are proper constants of the motion.
 The solution (\ref{eq.14}) has an obvious cylindrical character: each given 
orbit, characterized by the phase $\theta_{0}$, slides on the cylinder that 
is the direct product of a circle with radius $\rho$ in the plane 
$y_{1},y_{3}$, with a hyperbola with parameter $\epsilon$ in the plane 
$y_{2},y_{4}$. Obviously, this topology is preserved in the original 
coordinates, due to the convergence of (\ref{eq.3}). For a linear flow 
generated just by $h_2$ in (\ref{eq.1.1}) all frequencies $\Omega$ and 
velocities $\Lambda$ degenerate respectively into $\omega$ and $\lambda$,
 no matter what particular cylinder is considered.

The suitable form of the solution (\ref{eq.14}) immediately reveals some 
important structures near the saddle point. Let us consider $\rho=0$, in
 which case the motion is confined to the plane $y_{2},y_{4}$. For 
$\epsilon=0$, we identify the coordinate axes $y_{2}$ and $y_{4}$ as just
 the stable and unstable orbits originating from the saddle point. For
 $\epsilon \neq 0$, we have hyperbolae near the axes.

Consider now the case $\rho \neq 0$ (see figure \ref{Figura2.1}). If
 $y_{2}=y_{4}=0$ ($\epsilon=0$), we have the family of unstable (circular) 
periodic orbits confined to the plane $y_{1},y_{3}$. Let $\tau$ be one such 
orbit. If now only $y_{4}=0$, then the motion is composed by $\tau$ in that 
plane and the axis $y_{2}$, thus generating from $\tau$ the two unstable 
cylinders ($U_{1}$ and $U_{2}$). On the other hand, if only $y_{2}=0$, the 
composition of $\tau$ with the axis $y_4$ generates the pair of stable 
semicylinders ($S_{1}$ and $S_{2}$). Now, let us intersect any of the four 
semicylinders above with a plane transverse to it. As a consequence of the 
Poincar\'e-Cartan theorem \cite{18}, the symplectic area or action of the 
closed irreducible curve, formed at the intersection, will equal that of 
$\tau$. In fact, this is valid for any irreducible curve over the 
semicylinder. On the other hand, it is a lagrangian surface i.e. all 
reducible curves on it have null action. Finally, if we compose a circular 
motion in the plane $y_{1},y_{3}$ with a hyperbola in the plane $y_{2},y_{4}$ 
($\epsilon \neq 0$), we do not have semicylinders but smooth cylinders near 
the saddle point, distinct from those associated to the orbits $\tau$.

Of course, the ``rectified'' structures described above in the $\underline
{y}$ coordinates, will appear distorted --- yet preserving their topology 
--- in the original coordinates. Moreover, the distorsions are such that,
 far away from the saddle point, the cylinders execute a very complicated 
tangle among themselves if the regime is chaotic \cite{17}. Because they 
are all confined to a compact energy surface in phase space, transverse 
crossings of such cylinders eventually occur far from the origin. The 
existence of those transverse crossings are the source of chaotic motion in 
the continuum flow. So, an orbit at a transverse intersection of an unstable
 semicylinder with a stable one ($U_{1}$ with $S_{2}$ or $U_{2}$ with $S_{1}$ 
in the discussion above) is just a homoclinic orbit which tends to the 
respective orbit $\tau$ as $t \rightarrow \pm\infty$. In the case of the 
families of smooth cylinders in the neighbourhood of ($U_{1},S_{2}$) or 
($U_{2},S_{1}$), the selfcrossings eventually lead to closed orbits with 
increasing periods, which accumulate on the homoclinic orbits. All these 
continuum structures underly the homoclinic tangle observed in a (convenient,
 yet generally nonplanar) Poincar\'e section of the flow.

Because of its convergence, the MNF formalism allows us to start from
 precise initial conditions near the saddle point. Moreover, the preceding 
discussion shows the necessity of the MNF for the general definition of a 
given cylinder in phase space. This permits us to extend the (un)stable 
cylinders beyond the vicinity of the origin~\cite{17} and proceed to an 
accurate quantitative study of the structures presented above.

\section{RECURRENCE RELATIONS}\label{sec.3}

To consistently determine the coefficients $X(i,\underline{\ell})$ of 
(\ref{eq.3}) and $K(2\underline{m})$ of (\ref{eq.7}), we insert (\ref{eq.3})
 in (\ref{eq.2}) and compare the result with the system directly obtained 
from the hamiltonian (\ref{eq.7}). We see that this task, simple in 
principle, is laborious if we are interested in an algorithm  to compute 
the coefficents $X(i,\underline{\ell})$ and  $K(2\underline{m})$, up to 
an arbitrary order. Anyway, the recurrence relations we obtain are the
 following:
\begin{eqnarray}
\lefteqn{A(1,\underline{n})\,X(1,\underline{n})+B(1,\underline{n})
\,K(2n_{1},2n_{2})\;=}\nonumber\\
&&(n_{3}+1)\,H(n_{1},n_{3}+1,n_{2},n_{4})+\nonumber\\
&&\Theta(n-3)\sum_{\ell=2}^{n-1}(\ell_{3}+1)\,H(\ell_{1},\ell_{3}+1,\ell_{2}
,\ell_{4})\,Z_{\underline{n}}^{\underline{\ell}}-\nonumber\\
&&\Theta(n-4)\sum_{m=2}^{INT(n/2)}K(2\underline{m})\,W_{1,\underline{n}}^
{\underline{m}}{\mbox{\hspace{0.4cm},}}\label{eq.15}
\end{eqnarray}
\newpage
\begin{eqnarray}
\lefteqn{A(3,\underline{n})\,X(3,\underline{n})+B(3,\underline{n})\,K
(2n_{3},2n_{2})\;=}\nonumber\\
&&-(n_{1}+1)\,H(n_{1}+1,n_{3},n_{2},n_{4})-\nonumber\\
&&\Theta(n-3)\sum_{\ell=2}^{n-1}(\ell_{1}+1)\,H(\ell_{1}+1,\ell_{3},\ell_{2}
,\ell_{4})\,Z_{\underline{n}}^{\underline{\ell}}-\nonumber\\
&&\Theta(n-4)\sum_{m=2}^{INT(n/2)}K(2\underline{m})\,W_{3,\underline{n}}^{
\underline{m}}{\mbox{\hspace{0.4cm},}}\label{eq.16}
\end{eqnarray}
\begin {eqnarray}
\lefteqn{A(2,\underline{n})\,X(2,\underline{n})+B(2,\underline{n})\,K(2n_{1}
,2n_{2})\;=}\nonumber\\
&&(n_{4}+1)\,H(n_{1},n_{3},n_{2},n_{4}+1)+\nonumber\\
&&\Theta(n-3)\sum_{\ell=2}^{n-1}(\ell_{4}+1)\,H(\ell_{1},\ell_{3},\ell_{2}
,\ell_{4}+1)\,Z_{\underline{n}}^{\underline{\ell}}-\nonumber\\
&&\Theta(n-4)\sum_{m=2}^{INT(n/2)}K(2\underline{m})\,W_{2,\underline{n}}^{
\underline{m}}{\mbox{\hspace{0.4cm},}}\label{eq.17}
\end{eqnarray} 
\begin{eqnarray}
\lefteqn{A(4,\underline{n})\,X(4,\underline{n})+B(4,\underline{n})\,K(2n_{1}
,2n_{4})\;=}\nonumber\\
&&-(n_{2}+1)\,H(n_{1},n_{3},n_{2}+1,n_{4})-\nonumber\\
&&\Theta(n-3)\sum_{\ell=2}^{n-1}(\ell_{2}+1)\,H(\ell_{1},\ell_{3},\ell_{2}+1,
\ell_{4})\,Z_{\underline{n}}^{\underline{\ell}}-\nonumber\\
&&\Theta(n-4)\,\sum_{m=2}^{INT(n/2)}K(2\underline{m})\,W_{4,\underline{n}}^{
\underline{m}}{\mbox{\hspace{0.4cm}.}}\label{eq.18}
\end{eqnarray}
The definitions involved here are:
\begin{displaymath}
INT(x)\;=\;integer\;part\;of\;x{\mbox{\hspace{0.4cm},}}
\end{displaymath}
\begin{displaymath}
\Theta(x)\;=\;1\;if\;x \geq 0\;\;or\;\;0\;if\;x < 
0{\mbox{\hspace{0.4cm},}}
\end{displaymath}
\begin{equation}
\left\{
\begin{array}{l}
A(1,\underline{n})\;=\;\lambda_{1}(n_{1}-n_{3}-1)+\lambda_2(n_{2}-n_{4}) \\
A(3,\underline{n})\;=\;\lambda_{1}(n_{1}-n_{3}+1)+\lambda_2(n_{2}-n_{4}) \\
A(2,\underline{n})\;=\;\lambda_{1}(n_{1}-n_{3})+\lambda_2(n_{2}-n_{4}-1) \\
A(4,\underline{n})\;=\;\lambda_{1}(n_{1}-n_{3})+\lambda_2(n_{2}-n_{4}+1)
{\mbox{\hspace{0.4cm},}}\label{eq.21}
\end{array}
\right.
\end{equation}
\begin{equation}
\left\{
\begin{array}{l}
B(1,\underline{n})\;=\;n_{1}\delta_{n_{3}}^{n_{1}-1}\delta_{n_{4}}^{n_{2}}\\
B(3,\underline{n})\;=\;-n_{3}\delta_{n_{3}}^{n_{1}+1}\delta_{n_{4}}
^{n_{2}} \\
B(2,\underline{n})\;=\;n_{2}\delta_{n_{3}}^{n_{1}}\delta_{n_{4}}^{n_{2}-1} \\
B(4,\underline{n})\;=\;-n_{4}\delta_{n_{3}}^{n_{1}}\delta_{n_{4}}^{n_{2}+1}
 {\mbox{\hspace{0.4cm},}}\label{eq.22}
\end{array}
\right.
\end{equation}
\begin{displaymath}
\delta_{j}^{i}\;=\;1\;if\;i=j\;\;or\;\;0\;if\;i \neq 
j{\mbox{\hspace{0.4cm},}}
\end{displaymath} 
\begin{eqnarray}
\lefteqn{W_{i,\underline{n}}^{\underline{m}}\;=\;m_{1}(n_{1}-n_{3})\,
X(i,n_{1}
-m_{1}+1,n_{3}-m_{1}+1,n_{2}-m_{2},n_{4}-m_{2})\times}\nonumber\\
&&\times\Theta(n_{1}-m_{1}+1)\,\Theta(n_{3}-m_{1}+1)\,\Theta(n_{2}-m_{2})\,
\Theta(n_{4}-m_{2})+\nonumber\\
&&m_{2}(n_{2}-n_{4})\,X(i,n_{1}-m_{1},n_{3}-m_{1},n_{2}-m_{2}+1,n_{4}
-m_{2}+1)\times\nonumber\\
&&\times\Theta(n_{1}-m_{1})\,\Theta(n_{3}-m_{1})\,\Theta(n_{2}-m_{2}+1)
\,\Theta(n_{4}-m_{2}+1){\mbox{\hspace{0.4cm}.}}\label{eq.24}
\end{eqnarray}

Given $\underline{n}$ and $\underline{m}$, the coefficients 
$W_{i,\underline{n}}^{\underline{m}}$ depend only on the $X(i,\underline{k})$ 
of order\\ $k=n-2m+2$. As the minimum value of $m$ is 2, it follows that $k 
\leq n-2$ and hence $k<n$. As the maximum value of $\underline{m}$ is
$INT(n/2)$, it also follows that $k\ge2$.

The coefficients $Z_{\underline{n}}^{\underline{\ell}}$ arise through
 the relations
\begin{eqnarray}
\lefteqn{\underline{x}^{\underline{\ell}}\;=\;[x_{1}(\underline{y})]^
{\ell_{1}}\,[x_{3}(\underline{y})]^{\ell_{3}}\,[x_{2}(\underline{y})]
^{\ell_{2}}\,[x_{4}(\underline{y})]^{\ell_{4}}}\nonumber\\
&&=\;[y_{1}+\sum_{i=2}^{\infty}X(1,\underline{i})\,\underline{y}
^{\underline{i}}]^{\ell_{1}}\,[y_{3}+\sum_{j=2}^{\infty}X(3,
\underline{j})\,\underline{y}^{\underline{j}}]^{\ell_{3}}\times\nonumber\\
&&\times[y_{2}+\sum_{k=2}^{\infty}X(2,\underline{k})\,\underline{y}
^{\underline{k}}]^{\ell_{2}}\,[y_{4}+\sum_{m=2}^{\infty}X(4,\underline{m})
\,\underline{y}^{\underline{m}}]^{\ell_{4}}\nonumber\\
&&\equiv\;\underline{y}^{\underline{\ell}}+\sum_{n=\ell+1}^{\infty}\,
Z_{\underline{n}}^{\underline{\ell}}\,\underline{y}^{\underline{n}}
{\mbox{\hspace{0.4cm}.}}\label{eq.25}
\end{eqnarray}

Obtaining the coefficients $Z_{\underline{n}}^{\underline{\ell}}$ is the
 heaviest computational task of the whole algorithm. The reason, simply
stated, is that we have to obtain a series resulting from multiplying four 
terms, each of them being itself a series powered to an arbitrary integer!
 Given an index vector $\underline{\ell}$, the coefficients 
$Z_{\underline{n}}^{\underline{\ell}}$ depend only on the $X(i,\underline{k})
$ with $k < n$. To obtain the $Z_{\underline{n}}^{\underline{\ell}}$ up to 
order $n=N$, the series in (\ref{eq.25}) needs to be truncated at combined 
minimal orders which depend on $N$. If $N$ changes (increases), these series'
 minimal truncations also change, modifying the values of various 
$Z_{\underline{n}}^{\underline{\ell}}$ of order $n \le N$ previously 
obtained. Then, for each increase of $N$, we need to compute {\it{all}} 
$Z_{\underline{n}}^{\underline{\ell}}$ again, since we do not know (at least 
at the present stage) what coefficients, coming from the previous step, will 
not be changed by the actual calculation. Nevertheless, the process evidently 
converges as $N \rightarrow \infty$.

Now, let us consider a subset of coefficients 
$Z_{\underline{n}}^{\underline{\ell}}$ characterized by having some null 
indexes in $\underline{n} = (n_{1},n_{3},n_{2},n_{4})$. It is easy to see
 that this subset depends only on the $X(i,\underline{k})$ with 
$\underline{k}$ having the same null structure as $\underline{n}$. Then,
 the more null indexes the vector $\underline{n}$ has, the easier we can 
compute the corresponding subset $Z_{\underline{n}}^{\underline{\ell}}$ from 
the smaller number of coefficients of the respective subset 
$X(i,\underline{k})$. Now, let us return to the cylindrical structures 
discussed in the end of the last section. We need only the coefficients 
$X(i,n_{1},n_{3},0,0)$ or $X(i,0,0,n_{2},n_{4})$ to respectively describe,
 in the original coordinates, the normalized planes $y_{2}=y_{4}=0$ or
$y_{1}=y_{3}=0$. Unfortunately, the coefficients $X(i,n_{1},n_{3},0,0)$ and 
$X(i,0,0,n_{2},n_{4})$ depend by themselves on the remaining ones, through 
their relations with the coefficients $K(2\underline{m})$. So, we find that 
the decoupling occurs only if just one index is nonnull, as for e. g. the 
subsets $X(i,0,0,n_{2},0)$ or $X(i,0,0,0,n_{4})$. In particular, the later 
coefficients are just those we need to compute the (un)stable manifolds 
originated at the saddle point. In that case, the corresponding coefficients 
$Z_{\underline{n}}^{\underline{\ell}}$ can be obtained up to a much higher 
order than the full set, as we will see in section \ref{sec.5}.

Let us check the consistency of the recurrence relations (\ref{eq.15}) to 
(\ref{eq.18}). Each first member is a linear combination of the unknowns 
$X(i,\underline{n})$ and $K(2\underline{m})$, and the second members depend
 on the known hamiltonian coefficients $H(\underline{\ell})$, on the 
$X(i,\underline{n})$ through $Z_{\underline{n}}^{\underline{\ell}}$ and 
$W_{i,\underline{n}}^{\underline{m}}$ and explicitly on the 
$K(2\underline{m})$. Nevertheless, the appearance of $X(i,\underline{n})$ 
and $K(2\underline{m})$ in the second members occurs at lower orders than 
in the corresponding first ones, as expected. Additionally, they have the 
following property: if $A(i,\underline{n}) \neq 0$ then $B(i,\underline{n})
=0$ and reciprocally. This is due to the eigenvalues $\lambda_{1}$ and 
$\lambda_{2}$ being independent over the reals. So, those relations naturally 
split into two cases: firstly, we have $A(i,\underline{n}) \neq 0$ and 
$B(i,\underline{n})=0$. This case allows us to obtain the corresponding 
$X(i,\underline{n})$ in a direct way.

The case $A(i,\underline{n})=0$ and $B(i,\underline{n}) \neq 0$ can only 
occurs for odd orders $n$. This case has redundancies so, in order to obtain 
all the $K(2\underline{m})$, it is only necessary to work with either the 
pair of eqs. (\ref{eq.15}) and (\ref{eq.17}) or (\ref{eq.16}) and 
(\ref{eq.18}). The coefficients $X(i,\underline{n})$ that remain undetermined 
in this case are the following: $X(1,k_{1}+1,k_{1},k_{2},k_{2})$, 
$X(3,k_{1},k_{1}+1,k_{2},k_{2})$, $X(2,k_{1},k_{1},k_{2}+1,k_{2})$ and 
$X(4,k_{1},k_{1},k_{2},k_{2}+1)$, with $k_{1}+k_{2}=1,2,3,\cdots$. In fact, 
the Moser theorem imposes some additional normalizing conditions to these 
``weak'' coefficients, which can be simply met by requiring that all of them 
vanish. This will be assumed here. It should be noted that requiring the 
coordinate transformation to be canonical reduces this freedom and may not be 
compatible with the conditions of Moser's theorem.

\section{PREPARATION OF THE H\'ENON-HEILES HAMILTONIAN}\label{sec.4}

We apply the preceding formalism to the H\'enon-Heiles hamiltonian:
\begin{equation}
h(q_{1},q_{2},p_{1},p_{2})\;=\;\frac{1}{2}(p_{1}^{2}+p_{2}^{2})+U(q_{1}
,q_{2}){\mbox{\hspace{0.4cm},}}\label{eq.3.1}
\end{equation}
\begin{equation}
U(q_{1},q_{2})\;=\;\frac{1}{2}(q_{1}^{2}+q_{2}^{2}+2q_{1}^{2}q_{2}-\frac{2}
{3}q_{2}^{3}){\mbox{\hspace{0.4cm}.}}\label{eq.3.0}
\end{equation}
Here, $q_{1}$ and $q_{2}$ are the positions and $p_{1}$ and $p_{2}$ are
 their respective momenta.

Even though the motivation for the hamiltonian (\ref{eq.3.1}) was 
astronomical, it soon became relevant on its own, due to both its simplicity 
and its dynamical richness \cite{20,5,10,21}. For, if the energy $h\equiv E$ 
is suficiently small, the chaotic regions of the flow are vanishingly small. 
On the other hand, if the energy increases, the chaotic regions arbitrarily 
increase too. Additionally, (\ref{eq.3.1}) comes from the truncation of the 
Toda hamiltonian (see e.\ g. ~\cite{10}), the later one being  totally 
integrable for all energies.

In figure \ref{Figura3.1} we exhibit the equipotential curves of 
(\ref{eq.3.0}). The origin $P_{0}(q_1=0,p_1=0,q_2=0,p_2=0)$ is an eliptic 
(stable) equilibrium point with the eigenvalues $\pm i$, while 
$P_{1}(0,0,1,0)$, $P_{2}(-\sqrt{3}/2,0,-1/2,0)$ and 
$P_{3}(\sqrt{3}/2,0,-1/2,0)$ are saddle (unstable) points, all having the 
same set of eigenvalues $\pm i\sqrt{3},\pm 1$. These saddle points are just
 as required by our method. The evident trigonal symmetry of figure 
\ref{Figura3.1} makes them mutually equivalent from a dynamical point of
view. Then, there is just one MNF valid around $P_{1}$, $P_{2}$ and $P_{3}$.

To put the hamiltonian (\ref{eq.3.1}) in the suitable form (\ref{eq.1}), we 
need to perform the following coordinate transformations for 
$P_{j}\;,\;j=1,2,3$:
\begin{equation}
\left(
\begin{array}{l}
q_{1}\\
p_{1}\\
q_{2}\\
p_{2}
\end{array}
\right)
\;=\;
\left(
\begin{array}{l}
\tilde{q}_{1}\\
\tilde{p}_{1}\\
\tilde{q}_{2}\\
\tilde{p}_{2}
\end{array}
\right)
\;+\;
P_{j}^{T}{\mbox{\hspace{0.4cm},}}\label{eq.3.5}
\end{equation}
\begin{equation}
\left(
\begin{array}{l}
\tilde{q}_{1}\\
\tilde{p}_{1}\\
\tilde{q}_{2}\\
\tilde{p}_{2}
\end{array}
\right)
\;=\;
\underline{A}_{j}\;
\left(
\begin{array}{l}
\tilde{z}_{2}\\
\tilde{z}_{1}\\
\tilde{z}_{4}\\
\tilde{z}_{3}
\end{array}
\right){\mbox{\hspace{0.4cm},}}\label{eq.3.6}
\end{equation}
\begin{equation}
\left(
\begin{array}{l}
\tilde{z}_{1}\\
\tilde{z}_{3}\\
\tilde{z}_{2}\\
\tilde{z}_{4}
\end{array}
\right)
\;=\;
\left(
\begin{array}{cccc}
\frac{1}{\sqrt[4]{3}} & 0 & 0 & 0 \\
0 & \sqrt[4]{3} & 0 & 0 \\
0 & 0 & 1 & 0 \\
0 & 0 & 0 & 1 \\
\end{array}
\right)\;
\left(
\begin{array}{l}
z_{1}\\
z_{3}\\
z_{2}\\
z_{4}
\end{array}
\right){\mbox{\hspace{0.4cm},}}\label{eq.3.7}
\end{equation}
\begin{equation}
\left(
\begin{array}{l}
z_{1}\\
z_{3}\\
z_{2}\\
z_{4}
\end{array}
\right)
\;=\;
\frac{1}{\sqrt{2}}\;
\left(
\begin{array}{cccc}
1 & i & 0 & 0 \\
i & 1 & 0 & 0 \\
0 & 0 & 1 & 1 \\
0 & 0 & -1 & 1
\end{array}
\right)\;
\left(
\begin{array}{l}
x_{1}\\
x_{3}\\
x_{2}\\
x_{4}
\end{array}
\right){\mbox{\hspace{0.4cm}.}}\label{eq.3.8}
\end{equation}
Here, $P_{j}^{T}$ is the transpose of the respective saddle points 
given in the text and the matrices $\underline{A}_{j}$ are given by
\begin{equation}
\underline{A}_{1}
\;=\;
\left(
\begin{array}{cccc}
0 & 1 & 0 & 0 \\
1 & 0 & 0 & 0 \\
0 & 0 & 0 & 1 \\
0 & 0 & 1 & 0
\end{array}
\right)\label{eq.3.9}
\end{equation}
if the point is $P_{1}$, and
\begin{equation}
\underline{A}_{j}
\;=\;
\left(
\begin{array}{cccc}
cos\theta_{j} & sin\theta_{j} & 0 & 0 \\
-sin\theta_{j} & cos\theta_{j} & 0 & 0 \\
0 & 0 & cos\theta_{j} & sin\theta_{j} \\
0 & 0 & -sin\theta_{j} & cos\theta_{j}
\end{array}
\right),{\mbox{\hspace{0.4cm}$j=2,3$\hspace{0.4cm},}}\label{eq.3.10}
\end{equation}
if the point is \,$P_{2}$ or $P_{3}$, with $\theta_{2}=\frac{5\pi}{6}$ and
 $\theta_{3}=\frac{\pi}{6}$. All transformations made here are symplectic.

In the intermediate $\underline{z}$ coordinates, the hamiltonian form is
\begin{equation}
h(\underline{z})\;=\;\frac{1}{2}(z_{4}^{2}-z_{2}^{2})+\frac{\sqrt{3}}{2}
(z_{3}^{2}+z_{1}^{2})+\frac{\sqrt{3}}{3}z_{1}^{2}z_{2}-\frac{1}{3}z_{2}^
{3}+\frac{1}{6}{\mbox{\hspace{0.4cm}.}}\label{eq.3.11.a}
\end{equation}
The dissociation energy $E=1/6$ arises explicitly here. The final hamiltonian
 form, in the $\underline{x}$ coordinates, is

\begin{equation}
h(x_{1},x_{3},x_{2},x_{4})\;=\;i\sqrt{3}x_{1}x_{3}-x_{2}x_{4}+\frac{\sqrt{6}}
{12}(x_{1}+ix_{3})^{2}(x_{2}+x_{4})-\frac{\sqrt{2}}{12}(x_{2}+x_{4})^{3}
{\mbox{\hspace{0.4cm},}}\label{eq.3.11}
\end{equation}
where we have dropped the constant energy term. We obtain the coefficients 
$\lambda_{1},\lambda_{2}$ and $H(\underline{\ell})$ by identification of 
(\ref{eq.3.11}) with (\ref{eq.1}). We see that, for each $P_{j}$, we arrive
 at the same prepared hamiltonian form (\ref{eq.3.11}) and hence we have
 a single MNF valid for all three $P_{j}$, as expected.

\section{RESULTS}\label{sec.5}

To present the results for the H\'enon-Heiles hamiltonian, we will use the 
$\underline{z}$ coordinates of the last section. All numerical propagations 
were made via an optimized fourth order Runge-Kutta code \cite{22} and, for 
the Poincar\'e sections, we use a trick by H\'enon \cite{23} for exact 
crossings of the surface sections. 

First of all, we compute the coefficients of the special form 
$X(i,0,0,n_{2},0)$ and $X(i,0,0,0,n_{4})$. As we saw in section \ref{sec.3}, 
this computation is much easier than for the full set $X(i,\underline{n})$.
 So, with the same machinery and twice the real time for the general case, we
arrive only at the order 16 for the full set, against the order 50 for both 
subsets. The maximum order obtained for the coeficients $K(2\underline{m})$ 
was also 16.

Figures \ref{Figura3.4.a} and \ref{Figura3.4.b} show both the exact 
(un)stable manifolds that originate at the saddle point and their 
approximations via the coordinate transformations 
$\underline{z}(\underline{y})$ (see (\ref{eq.3.8}) and (\ref{eq.3})), 
truncated at various orders up to the maximum of 50. We see that the exact 
curves meet smoothly in the bounded region. In fact, they pertain to a family 
of orbits entirely confined to the plane $z_{1}=z_{3}=0$, as is easily seen 
from the form (\ref{eq.3.11.a}) of the H\'enon-Heiles hamiltonian. To obtain 
the exact curves, we select one starting point near the origin at each axis 
$y_{2}$ or $y_{4}$ (see (\ref{eq.14})) and then propagate them through the 
exact flow generated by $h(\underline{z})$. This method will be referred to
 in the following as {\it{z-propagation}}, being always precise, provided
 we start from $\underline{y}$ near the origin. The z-propagation is 
semianalytical in the sense that we numerically evolve precise initial
conditions that are {\it{analytically}} given. We note that for localizing
 the ``correct'' initial directions i.e. those given by the axes $y_{2}$
 and $y_{4}$, the form of (\ref{eq.14}) is crucial. On the other hand,
 the approximate curves in figures \ref{Figura3.4.a} and \ref{Figura3.4.b}
 are just the axes $y_{2}$ and $y_{4}$, merely rewritten in the coordinates 
$\underline{z}$ via the series truncated at different orders. This last
 method will be referred to in the following as {\it{yz-translation}} and
 its precision is only guaranted by Moser's theorem if $\underline{y}$
 remains in the neighbourhood of the origin. The yz-translation is 
analytical in the sense that it uses only algebraic calculations
 and properties of the normal form.

The remarkable feature of both figures is that the analytic yz-translation 
clearly points to the convergence of the series (\ref{eq.3}) even {\it{far 
from}} the small region where Moser initially guaranteed it. This (and 
additional evidence in the following) strongly suggests the possibility of 
extending Moser's original convergence region. This issue is being considered 
in our present research. We also note that the region of accurate 
approximation ceases more and more abruptly as the truncation order become 
higher.

The next task is to obtain the unstable periodic orbits near the saddle 
point. They are entirely confined to the plane $y_{2}=y_{4}=0$ ($\epsilon = 
0$) of the normalized system, the family parameter being $\rho$ (see 
(\ref{eq.14})). We need in principle only the coefficients of the form 
$X(i,n_{1},n_{3},0,0)$ to compute them. However, we saw in section
 \ref{sec.3} that those and all remaining coefficients must jointly
 be found. We obtained the full set $X(i,\underline{n})$ in this work 
until the order 16 and from now on this will be ever the case.

Figures \ref{Figura3.5.b} and \ref{Figura3.5.a} show the projections of
 three such unstable closed orbits in the planes $z_{2}=z_{4}=0$ and 
$z_{1}=z_{3}=0$, respectively. The values of $\rho$ caracterizing each orbit 
are shown in the figures. For sufficiently small values of $\rho$ we are 
always in the convergence region for $y_2 = y_4 = 0$ and both methods i.e. 
the z-propagation and yz-translation, generate closed curves which are 
indistinguishable in practice. Any orbit of the family closes once in figure 
\ref{Figura3.5.b} for every two returns in figure \ref{Figura3.5.a}. We also 
note that the motion is microscopic or ``residual'' in the second figure (in 
the plane $z_{1}=z_{3}=0$) relatively to the first one (in the plane 
$z_{2}=z_{4}=0$). These features are easily understood if we consider 
$y_2=y_4=0$ in (\ref{eq.3}) and (\ref{eq.3.8}) and keep only contributions 
from the $\ell = 2$ terms in (\ref{eq.3}).

We saw in section \ref{sec.3} that there exist a cylindrical flow associated 
to any of the unstable periodic orbits above. Four semicylinders emanate
 from each of them. Let $\tau$ be one such orbit and consider, for example, 
only the pair of (un)stable semicylinders which directly evolve from $\tau$
 to the region $z_{2} < 0$. This evolution occurs in an intrincate manner in 
the energy shell and, because the dynamics are chaotic, the semicylinders 
mutually intersect transversally. At each of these intersections there
 exists a homoclinic orbit (HO) which tends to $\tau$ as $t \rightarrow 
\pm\infty$.

We consider in this work only the first of the intersections above, in
 which we have the {\it{primary}} HOs. A suitable Poincar\'e surface to 
observe that is one that is transverse to the cylinders themselves, i.e.,
 one that intersects each cylinder in an irreducible (closed) curve. The 
symmetry of figures \ref{Figura3.4.a} and \ref{Figura3.4.b} and the 
continuity of the cylinders with respect to the parameter $\rho$ suggest 
that one such surface is the plane $z_{4} = 0$. On the other hand, 
figure \ref{Figura3.5.a} shows that this plane also intersects the 
corresponding orbit $\tau$ ( and hence the semicylinders themselves)
 in the region $z_{2} > 0$. In fact, it is seen both qualitatively and 
numerically that in the region $z_{2} > 0$ the cylinders 
do not mutually intersect at all, except obviously at $\tau$ itself. 
Moreover, the plane $z_{4}=0$ generates only reducible (open) curves on the 
cylinders in that region i.e. it is longitudinal to the cylinders there. For
 sufficiently small $\rho$ the orbit $\tau$ is near the origin, so we avoid
numerical troubles, without losing precision, by starting on the cylinders 
at some small $z_{2} < 0$. 

We obtain the Poincar\'e section $z_{4} = 0$ in the region $z_{2} < 0$ by
three methods. The first one is the z-propagation of the following initial 
conditions: a sufficiently small (constant) height $y_{4} = y_{40}$
 determines a straight section on the semicylinder characterized by $y_{2}
 = 0$ (hence $\epsilon = 0$) and a given $\rho$ (see (\ref{eq.14})). Now,
 all orbits through the section are given by
\begin{equation}
\left\{
\begin{array}{l}
y_{1}\;=\;\rho\,\exp{[i\theta]} \\
y_{3}\;=\;-i\rho\,\exp{[-i\theta]}{\mbox{\hspace{0.4cm},\hspace{0.4cm}$\;
\theta\;\in\;[0,2\pi)$\hspace{0.4cm},}}\label{eq.3.14}
\end{array}
\right.
\end{equation}
where the phase $\theta$ identifies each orbit on it. Obviously, we
 translate all initial points above from $\underline{y}$ to $\underline{z}$, 
before propagating them by the exact flow. The propagation of the other 
semicylinder, characterized by $y_{4} = 0$ and the same $\rho$, is made
 in a similar way, except that the straight section is now determined 
by the small height $y_{2} = y_{20}$. In all cases presented here it
 proved to be sufficient to use $y_{20} = y_{40} = 1.0\times10^{-2}$. We 
z-propagated both semicylinders until the condition $z_{4} = 0$ was 
reached by the first time in the region $z_{2} < 0$. 

Figure \ref{Figura3.9.a} shows a typical Poincar\'e section as described 
above, namely that for $\rho = 1.0\times10^{-3}$. Four primary HOs are 
exhibited there. The highly symmetric features of this figure (and all 
succeeding ones) must be viewed as reflecting the intrinsic symmetries of
 the H\'enon-Heiles potential. So, they do not depend on the specific value 
of $\rho$, which works somewhat as a scaling factor in all them. In 
particular, we see that the HOs are always on the axes, as indicated in 
figure \ref{Figura3.9.a}. Each of these HOs emanates from the corresponding 
orbit $\tau$ along the unstable cylinder and returns to $\tau$ along the
 stable one. The remaining orbits do not exchange cylinders, at least at 
the first crossing. Figures \ref{Figura3.11.a}, \ref{Figura3.11.b}, 
\ref{Figura3.11.c} and \ref{Figura3.11.d} show, for $\rho = 
1.0\times10^{-3}$, the evolution of a typical orbit on each cylinder until
 the Poincar\'e section was reached. On the other hand, figures 
\ref{Figura3.18}, \ref{Figura3.30} and \ref{Figura3.19} show a primary HO 
associated to the orbit $\tau$ with $\rho = 2.6\times10^{-3}$, just that
 which corresponds to the HO localized on the left (i.e. at $z_{3} < 0$) 
in figure \ref{Figura3.9.a}. Figure \ref{Figura3.30} is an amplification 
of figure \ref{Figura3.18}, revealing the microstructure of the HO near 
$\tau$. We note that two typical orbits on distinct semicylinders do not 
have the same coordinates when they reach, for the first time, the condition 
$z_{4} = 0$, contrary to what occurs with the two branches of a primary 
homoclinic curve.

We also consider two analytical methods for the determination of the same
 Poincar\'e sections above. One of them is the yz-translation already 
presented. For this purpose we have merely to rewrite the condition $z_{4}
 = 0$ in the $\underline{y}$ coordinates,
\begin{equation}
y_{4}^{N+1}+\sum_{n=2}^{N}[X(2,n_{1},n_{3},0,n_{4})-X(4,n_{1},n_{3},0,
n_{4})]y_{1}^{n_{1}}y_{3}^{n_{3}}y_{4}^{N+n_{4}}\;=\;0{\mbox{\hspace{0.4cm}
,}}\label{eq.3.13}
\end{equation}
for the semicylinder for which $y_{2}=0$ and
\begin{equation}
y_{2}^{N+1}+\sum_{n=2}^{N}[X(2,n_{1},n_{3},n_{2},0)-X(4,n_{1},n_{3},
n_{2},0)]y_{1}^{n_{1}}y_{3}^{n_{3}}y_{2}^{N+n_{2}}\;=\;0{\mbox{\hspace{0.4cm}
,}}\label{eq.3.12}
\end{equation}
for the one with $y_{4}=0$. Here, $N = 16$ and $y_{1}$ and $y_{3}$ are
 fixed through (\ref{eq.3.14}) for each value of $\theta$. The equations 
(\ref{eq.3.13}) and (\ref{eq.3.12}) thus determine for each  phase/orbit 
$\theta$ what is the corresponding height $y_{4}$ or $y_{2}$ where it
 crosses the plane $z_{4} = 0$. As they are polynomials of degree $2N$ 
in $y_{4}$ and $y_{2}$ respectively, we have to select in each case, 
among the $2N$ complex roots, the only one that is {\it{real}} and verifies 
$z_{2} < 0$. To find the roots, we use codes described in \cite{22}, based
 on the Laguerre method. Finally, we come back to the results in the
$\underline{z}$ coordinates.

The second analytical method is essentially equivalent to the 
yz-translation and consists of evolving in time the solutions
 (\ref{eq.14}) themselves until the condition $z_{4} = 0$ be verified.
 We will refer to it as the {\it{y-propagation}}. Even though the 
y-propagation be trivial, the comparison of both analytical methods is an
important test for the numerical consistency of both sets of coefficients 
$X(i,\underline{n})$ and $K(2\underline{m})$ that we obtained, as the 
yz-translation involves {\it{only}} the coefficients $X(i,\underline{n})$.
 In fact, the figures for the Poincar\'e sections obtained through both 
analytical methods are indistinguishable in practice, which confirms the 
mutual consistency of those two sets of coefficients. On the other hand,
 in figure \ref{Figura3.10.a} we superimpose (for $\rho = 1.0\times10^{-3}$) 
the analytical and the semianalytical results, the latter already displayed 
in figure \ref{Figura3.9.a}. We conclude that the analytical method is 
able to describe all qualitative aspects, such as symmetries, and the 
existence of the transverse cylindrical intersections and HOs --- and hence 
the chaotic nature of the motion. Moreover, we see that even its numerical 
efficiency is surprising if we realize that we are far from the region of 
guaranteed convergence. Considering that here the MNF has only been expanded 
to order 16, we are faced again with an indication that Moser's theorem can 
be extended.

Now we are ready to search for the closed orbits with long periods, which 
accumulate on the HOs. For this purpose, we have to consider the smooth 
selfintersecting cylinders for which not only $\rho \neq 0$ but also 
$\epsilon \neq 0$ (see section \ref{sec.3}). Due to the continuity of the
cylindrical structure with respect to the parameters $\rho$ and $\epsilon$, 
we know that the transverse cylindrical selfcrossings must also occur at 
the plane $z_{4} = 0$ in the region $z_{2} < 0$. Then, all the previous 
discussion about both the semianalytical and analytical methods are still 
in order here, the only changes being now that the relation $y_{2}y_{4} = 
\epsilon$ eliminates both conditions  (\ref{eq.3.13}) and (\ref{eq.3.12}) 
in favour of just one, e.\ g.
\begin{equation}
y_{2}^{N+1}-\epsilon\,y_{2}^{N-1}+\sum_{n=2}^{N}[X(2,\underline{n})-X(4,
\underline{n})]y_{1}^{n_{1}}y_{3}^{n_{3}}\epsilon^{n_{4}}y_{2}^{N+n_{2}
-n_{4}}\;=\;0{\mbox{\hspace{0.4cm},}}\label{eq.3.15}
\end{equation}
with $y_{1}$ and $y_{3}$ still given by (\ref{eq.3.14}).

Figure \ref{Figura3.15} shows the Poincar\'e section above, obtained
 through the semianalytical method for the cylinder with parameters 
$\rho = \epsilon = 1.0\times10^{-3}$. It exhibits four orbits in the
 first autointersection of the cylinder, with the same symmetries already 
seen. Figures \ref{Figura3.17.a} and \ref{Figura3.17.b} show a typical
 orbit on that cylinder, from the time it crosses the plane $z_{4} = 0$ 
through the branch of the cylinder which brings the orbit near to the
 origin (the {\it{S-branch}}), until the time it crosses that plane again, 
coming through the branch which takes it away from the origin 
({\it{U-branch}}). We compared with each other the two analytical methods 
(yz-translation and y-propagation), confirming also in this case the 
underlying numerical consistency of both sets of coefficients 
$X(i,\underline{n})$ and $K(2\underline{m})$. In figure \ref{Figura3.16}
 we superimpose the analytical results with those displayed in figure 
\ref{Figura3.15}.

What happens to the orbits in the selfintersections of figure 
\ref{Figura3.15}? We easily discover by numerical computation that,
 they do not close in general for any specific values of $\rho$ and 
$\epsilon$ --- at least at the {\it{first}} intersection. In fact, 
as the S and U-branches of the cylinder met smoothly near the origin, 
a more complex evolution can occur to {\it{any}} orbit before it
eventually closes after a certain number of cylindrical selfintersections.
 We consider here the accumulation of periodic orbits only on the primary 
HOs, so we will limit the search of closed orbits just at the first crossing.
 For this reason, we fix $\rho$ at some value and vary $\epsilon$, thus 
verifying for each $\epsilon$ whether the four primary autointersections 
close on themselves or not. They in fact do so for a set of well defined 
discrete values of $\epsilon$. In particular, the four curves close 
simultaneously on themselves when they do so, generating four distinct 
periodic orbits of the same period. We display in table \ref{tabela1} the
 numerical values of energy and period for the case $\rho = 
2.6\times10^{-3}$. We present there only those orbits at the crossing 
localized on the left (with $z_{3} < 0$) in figure \ref{Figura3.15}, while
 the corresponding primary HO of interest lies at the same crossing, as in 
figure \ref{Figura3.9.a}. PO1,...,PO4 are the periodic orbits we found,
 E is the energy and T is the period of each orbit. For the HO the finite 
numerical value for T is also shown. Due to the high numerical instability 
associated to the chaotic regime, the orbit are quickly lost after that time.
 We note the fast convergence of the values of E and also that the periods
 of the POs are (nearly) multiple of that of $\tau$. On the other hand, the 
accumulation process of PO1,...,PO4 onto the corresponding HO, which in turn 
tends to $\tau$ as $t \rightarrow \pm \infty$, is clearly shown in figures 
\ref{Figura3.35}, \ref{Figura3.36} and \ref{Figura3.37}. These figures are 
typical of the behaviour of the periodic orbits for all other values of 
$\rho$.

Finally, an independent test was made to assure that the periodic orbits
 above are computed with precision by the present method, i.e. we ask
 whether the truncation at $N=16$ of the series (\ref{eq.3}) is sufficient
 or not to yield precise results. For this sake we use a code \cite{24} 
already known from the literature for computing periodic orbits in
 hamiltonian flows. By starting from a guessed periodic orbit, the code 
searches for an exactly closed orbit in its vicinity (in terms of either
 the energy or the period).  That code is based on rewriting the linear
finite difference integration of the flow in terms of the monodromy matrix 
and its main feature is the very fast convergence rate.

So, we enter as guessed periodic orbits, the closed ones we have obtained 
above and compare each of them with the corresponding exact curve we get
 through the code. It turns out that both orbits are indistinguishable
 in all cases we considered. For example, the exact curves fit completely 
those shown in figures \ref{Figura3.35} to \ref{Figura3.37}, at those
 scales. Hence, we conclude that the truncation at $N=16$ we have used is 
sufficient to support the semianalytical results presented here.

\section{CONCLUSIONS}\label{sec.6}

We have applied the theory of Moser's convergent normal forms to a
 two-degree of freedom hamiltonian flow near a saddle point. It turns
 out that  the cylindrical topology of the flow is promptly revealed 
in the normal coordinates. The chaotic behaviour was studied directly
in the continuous flow underlying the homoclinic tangle in a Poincar\'e 
section. We applied the method to the H\'enon-Heiles hamiltonian.
This allowed to compute structures related to chaotic motion in the full
 phase space rather than just a Poincare\'e section: the (un)stable
 manilfolds originated at the saddle point, the unstable periodic orbits 
with the cylinders they generate and the homoclinic orbits associated to 
them, and also the closed orbits with long periods which accumulate on
 the homoclinic ones.

We developed two methods for obtaining those structures. The first one is 
semianalytical in the sense that we numerically propagate initial conditions 
that we were able to find only through the normal form. The second method
 (in two versions --- which are important to test the numerical consistency 
of all coefficients obtained) is entirely analytical, in the sense that it 
uses only algebraic calculations and properties of the normal form. The 
calculations are truncated at the order 16 and compared with a known
 numerical method for searching exact periodic orbits in a two-degree 
hamiltonian flow. Our semianalytical results proved accurate, thus showing
 the sufficiency of the truncation order that we have used. On the other 
hand, the analytical method was surprisingly good even far away from the 
saddle point and it was certainly able to exhibit all the qualitative 
aspects of the chaotic motion there.

It was also possible to compute analytically, as a particular case, the 
(un)stable manifolds originated at the saddle point up to the order 50,
 obtaining much better numerical agreement for the analytical method. Both 
facts indicate the convergence of the normal form beyond the ``microscopic''
 region originally established by the Moser's theorem. We are presently 
working out this extension.

Despite the mathematical interest in ascertaining the convergence of
 Moser's normal form as far as the first intersection of the stable and 
unstable manifolds, there is no doubt that, even for a simple polynomial 
potential as the H\'enon-Heiles, the convergence will be very slow. In 
view of the huge difficulty in increasing the order of the normal form,
 it seems that the semianalytical method that we have used to obtain 
chaotic structures is to be preferred. This relies only on a relatively
 small truncation of the normal form near the saddle point, which produces 
accurate cylinders that may be extended by numerical integration.

\section{ACKNOWLEDGEMENTS}

The authors are grateful to Professors Marcus A. M. de Aguiar,
 Kyoko Furuya and Roberto F. S. Andrade for many discussions and
 suggestions, to Professor Sylvio F. Mello for the reading of the Ph.\ D. 
thesis (IFGW-UNICAMP) which originated this work, and to {\bf{CNPq}} for
 the financial support of this research.

\newpage

\begin{table}[h]
\begin{tabular}{|c|c|c|c|}\hline
 & $\epsilon$ & E & T \\ \hline
PO1&$5.836022\times10^{-5}$&0.1666200153&14.026 \\ \hline
PO2&$4.1216\times10^{-8}$&0.166678334123&21.381 \\ \hline
PO3&$2.9118\times10^{-11}$&0.166678375310&28.536 \\ \hline
PO4&$2.057124\times10^{-14}$&0.166678375339&35.791 \\ \hline
HO&0&0.166678375339&($\infty$) 83.795 \\ \hline
$\tau$&0&0.166678375339&3.628 \\ \hline
\end{tabular}
\begin{center}
{TABLE \protect{\ref{tabela1}}}
\end{center}
\end{table}

\newpage

\begin{center}
{TABLE CAPTIONS}
\end{center}

\begin{table}[h]
\caption{Parameters of some orbits discussed in the text
 with $\rho=2.6\times10^{-3}$.}\label{tabela1}
\end{table}

\newpage

\begin{figure}[h]
\caption{Projections in the planes $y_2=y_4=0$ (a) and $y_1=y_3=0$ (b)
 of a hamiltonian flow near a saddle point.}\label{Figura2.1}
\end{figure}

\begin{figure}[h]
\caption{Equipotential curves of the H\'enon-Heiles potential shown in eq. 
(\protect{\ref{eq.3.0}}).}\label{Figura3.1}
\end{figure}

\begin{figure}[h]
\caption{Exact (un)stable manifolds that originate at the saddle point and 
their MNF approximations at various even truncation orders 2, $\cdots$, 50. 
The exact curves meet smoothly at $z_4=-1.5$ and a particle in the loop runs 
in the clockwise direction. Note the evidence of convergence of the MNF far
 from the origin.}\label{Figura3.4.a}
\end{figure}

\begin{figure}[h]
\caption{The same as in the figure \protect{\ref{Figura3.4.a}} except that
 the truncations are made at the various {\it{odd}} orders shown. Note in 
this case the ``out'' direction of the divergence of the approximate curves 
from the exact one.}\label{Figura3.4.b}
\end{figure}

\begin{figure}[h]
\caption{Projection in the plane $z_{2}=z_{4}=0$ of three unstable 
periodic orbits (for which $y_2=y_4=0$ and hence $\epsilon=0$) near the
 saddle point, for the values of $\rho$ indicated in the figure.}
\label{Figura3.5.b}
\end{figure}

\begin{figure}[h]
\caption{Projection in the plane $z_{1}=z_{3}=0$ of the orbits shown in 
the figure \protect{\ref{Figura3.5.b}}. Note that two loops here 
corresponds to one loop in the other projection. Note also the microscopic 
motion here, relatively to that in the plane 
$z_{2}=z_{4}=0$.}\label{Figura3.5.a}
\end{figure}

\begin{figure}[h]
\caption{Poincar\'e section at $z_{4} = 0$ (in the region $z_{2} < 0$) of
 the (un)stable semicylinders (for which $y_2y_4=\epsilon=0$), originated
 at the orbit $\tau$ with $\rho=1.0\times10^{-3}$. Note the four primary 
homoclinic orbits at the intersections on the axes.}\label{Figura3.9.a}
\end{figure}

\begin{figure}[h]
\caption{Projection of the (time-reversed) evolution of a typical orbit on
the stable semicylinder, from a small initial condition already in the 
region $z_{2}<0$, until the condition $z_{4}=0$ was reached. The parameters 
are $\rho=1.0\times10^{-3}$ and $y_2y_4=\epsilon=0$.}\label{Figura3.11.a}
\end{figure}

\begin{figure}[h]
\caption{The other projection of the orbit shown in the figure 
\protect{\ref{Figura3.11.a}}.}\label{Figura3.11.b}
\end{figure}

\begin{figure}[h]
\caption{Evolution of a typical orbit on the unstable semicylinder 
in the same conditions of the figure 
\protect{\ref{Figura3.11.a}}.}\label{Figura3.11.c}
\end{figure}

\begin{figure}[h]
\caption{The other projection of the case shown in the figure 
\protect{\ref{Figura3.11.c}}.}\label{Figura3.11.d}
\end{figure}

\begin{figure}[h]
\caption{Projection of a primary homoclinic orbit (the HO indicated 
in the text), associated to the orbit $\tau$ with $\rho = 2.6\times10^{-3}$. 
The computed time was $T=83.795$, much longer than the period $T=3.628$ of 
$\tau$, and its energy $E=0.166678375339$ is the same of $\tau$, as
 expected (see table \protect{\ref{tabela1}}).}\label{Figura3.18}
\end{figure}

\begin{figure}[h]
\caption{Amplification of the figure \protect{\ref{Figura3.18}} showing
 the microstructure of the HO near the orbit $\tau$.}\label{Figura3.30}
\end{figure}

\begin{figure}[h]
\caption{The other projection of the homoclinic orbit shown in figure 
\protect{\ref{Figura3.18}}, compared with the corresponding orbit 
$\tau$.}\label{Figura3.19}
\end{figure}

\begin{figure}[h]
\caption{Superimposition of the precise semianalytical results, already 
displayed in the figure \protect{\ref{Figura3.9.a}}, to that obtained by
 the analytical method. Note the qualitative agreement and even the
 surprising numerical proximity between them, although we are here very 
distant of the origin. This is an additional evidence for the convergence
 of the Moser normal form beyond that neighbourhood.}\label{Figura3.10.a}
\end{figure}

\begin{figure}[h]
\caption{Precise semianalytical obtaining of the Poincar\'e section at
 $z_{4} = 0$ (in the region $z_{2} < 0$) of the (smooth) cylinder  for
 which  $\rho=\epsilon=1.0\times10^{-3}$. The resemblance with the figure 
\protect{\ref{Figura3.9.a}} is due to the intrinsic simmetries of the 
H\'enon-Heiles potential.}\label{Figura3.15}
\end{figure}

\begin{figure}[h]
\caption{Projection of a typical orbit on the (smooth) cylinder for which 
$\rho=\epsilon=1.0\times10^{-3}$, from a small initial condition already
 in the region $z_{2}<0$, until the condition $z_{4}=0$ was reached in
 both time directions.}\label{Figura3.17.a}
\end{figure}

\begin{figure}[h]
\caption{The other projection of the orbit shown in the figure 
\protect{\ref{Figura3.17.a}}. Note that it closes on itself on the plane
 shown in that figure but not here.}\label{Figura3.17.b}
\end{figure}

\begin{figure}[h]
\caption{Superimposition of the results already displayed in the figure 
\protect{\ref{Figura3.15}}, to that obtained by the analytical method. 
The comments made in figure \protect{\ref{Figura3.10.a}}, about the
 comparison of the semianalytical and analytical methods, are in order
 here.}\label{Figura3.16}
\end{figure}

\begin{figure}[h]
\caption{Superimposition of the orbits of the table \protect{\ref{tabela1}},
 showing the accumulation of the periodic orbits (PO1 to PO4 in the table)
 onto the corresponding homoclinic orbit (HO), associated to the $\tau$
 orbit with $\rho=2.6\times10^{-3}$. There is no difference among them far 
from the origin at this scale.}\label{Figura3.35}
\end{figure}

\begin{figure}[h]
\caption{Amplification of the figure \protect{\ref{Figura3.35}} near the
 origin. The accumulation process is much more slow here, on the contrary 
that occurs in the far distant region. The periodic orbits PO1 and PO2 of
 the table \protect{\ref{tabela1}} does not reach this region. 
}\label{Figura3.36}
\end{figure}

\begin{figure}[h]
\caption{The same accumulation process of figure \protect{\ref{Figura3.35}} 
projected now onto the other plane. These results are typical of all others 
values of $\rho$ we have considered.}\label{Figura3.37}
\end{figure}

\end{document}